# Energy Consumption and Battery Aging Minimization Using a Q-learning Strategy for a Battery/Ultracapacitor Electric Vehicle

Bin Xu[1], Junzhe Shi[2,*], Sixu Li[3], Huayi Li[4], Zhe Wang[1]

1: Clemson University, Department of Automotive Engineering, 4 Research Dr., Greenville, SC, 29607, USA
2: University of California, Berkeley, Civil and Environmental Engineering Department, 760 Davis Hall, Berkeley, CA 94720, USA.
3: Hunan University, Department of Mechanical and Vehicle Engineering, Changsha, Hunan Province, 410082, China.
4: University of Michigan, Department of Mechanical Engineering, 2350 Hayward St, Ann Arbor, MI 48109, USA.
* Corresponding author at: University of California, Berkeley, Civil and Environmental Engineering Department, 760 Davis Hall, Berkeley, CA 94720, USA. E-mail address: junzhe@berkeley.edu (Junzhe Shi).

*Abstract* – Propulsion system electrification revolution has been undergoing in the automotive industry. The electrified propulsion system improves energy efficiency and reduces the dependence on fossil fuel. However, the batteries of electric vehicles experience degradation process during vehicle operation. Research considering both battery degradation and energy consumption in battery/ supercapacitor electric vehicles is still lacking. This study proposes a *Q*-learning-based strategy to minimize battery degradation and energy consumption. Besides *Q*-learning, two heuristic energy management methods are also proposed and optimized using Particle Swarm Optimization algorithm. A vehicle propulsion system model is first presented, where the severity factor battery degradation model is considered and experimentally validated with the help of Genetic Algorithm. In the results analysis, *Q*-learning is first explained with the optimal policy map after learning. Then, the result from a vehicle without ultracapacitor is used as the baseline, which is compared with the results from the vehicle with ultracapacitor using Q-learning, and two heuristic methods as the energy management strategies. At the learning and validation driving cycles, the results indicate that the *Q*-learning strategy slows down the battery degradation by 13-20% and increases the vehicle range by 1.5-2% compared with the baseline vehicle without ultracapacitor.

1. Introduction

In the US, the government stated that the CO2 emissions of passenger cars should decrease from 132g/km in 2016 to 88g/kg in 2025 [1]. The emission regulations push the technologies into the limits. In response to the strict regulation, the entire automotive industry is experiencing a propulsion system electrification revolution. The propulsion system electrification results in different levels of electrification, such as mild hybrid electric vehicles (HEV) [2], plug-in HEV [3], and electric vehicles (EV) [4].

The electrified propulsion system requires energy storage systems (ESS), which generally include lithium ion batteries [5], fuel cells [6] and supercapacitors [7]. Lithium ion battery stores the electricity in the lithium ion during charging and releases energy when lithium ions cross the membrane during discharging [5]. Among the three ESSs, lithium ion batteries have middle range energy density, middle range power density and low cost [8]. Fuel cells release energy via the reaction between hydrogen and oxygen. Among the three ESSs, fuel cells have high energy density, middle range power density, and high cost [8]. Supercapacitors store electricity by charging two parallel aligned plates. The charge and discharge power can be extremely large, which indicates high power density. However, its energy density is low, thus pure supercapacitor-powered HEVs or EVs are seldom seen and it is commonly accompanied by lithium ion batteries [9] or fuel cells [10]. Lithium ion battery is the most popular ESS thanks to its combined features of low cost and high energy density.

Most researches have been focused on the energy efficiency and mileage of EV [11], while battery degradation during vehicle operation does not receive equal attention. It is reported that energy consumption and indirect greenhouse gas release increase because of battery degradation during vehicle operation [1]. In addition, experimental data from US Argonne National Lab showed that battery life only lasts less than 5 years in some states (assuming 30% capacity drop results in end of life) [1]. That could be part of the reasons the resale value of EVs are extremely low compared with engine only vehicles at similar mileage and class. Therefore, battery degradation during vehicle operation should be well investigated and minimized like the energy consumption.

There exist many great pure battery degradation studies without vehicle consideration. Choi et al. analyzed the degradation effect of LiCoO2 batteries based on experimental data [12]. High charge cutoff condition and high C-rate were found to cause severe degradation, whereas reducing depth of discharge did not show obvious degradation reduction. Xu et al. presented empirical battery degradation models, which were calibrated with experimental data [13]. The models were suitable for different types of batteries. In addition, battery degradation in vehicle can be found in some literature. Lam et al. proposed a practical capacity fading model for LiFePO4 battery in EV. The model was calibrated with experimental data under varying temperature, C-rate, state-of-charge (SOC) and depth of discharge with limited ranges. Zhang et al. considered supercapacitor in EV to mitigate the lithium ion battery degradation [14]. Rule-based energy management strategy (EMS) was redesigned for the supercapacitor/ battery EV. The results showed 30% less degradation with the addition of supercapacitor compared with pure battery ESS. Pelletier et al. reviewed the battery degradation for EV [15]. Battery capacity losses were plotted with respect to number of EV usage days at varying conditions, such as temperature, C-rate, SOC, and depth of discharge, which showed large impact on battery degradation. In existing EV literature, EMS mainly considered energy consumption minimization and battery degradation was not often taken into consideration [7, 9, 16-18]. In the literature considering battery degradation, many EMSs were rule-based methods [14, 15], which leave room for advanced EMS.

To overcome the aforementioned two research gaps, this study proposes a reinforcement learning (RL) based EMS and two optimized heuristic EMSs for the energy and battery aging optimization of battery/ ultracapacitor EV. There are few RL researches on the topic of pure EV EMS. Yang et al. implemented a Deep RL strategy in an ultracapacitor energy management for electric rail transits [19]. The Deep RL optimizes the maximum charge and discharge currents for the ultracapacitor. The proposed Deep RL strategy showed 2.65% energy saving than the rule-based strategy. However, the ultracapacitor in the electric rail application is much larger than that in passenger cars, thus the conclusion may not hold true in the passenger car application. Xiong et al. applied a tabular RL in s plug-in HEV energy consumption problem and found that the tabular RL consumed 16.8% less energy than a rule-based method. However, the battery aging is not included in the cost function and the rule-based method is not optimized. Sun et al. developed a RL EMS for a fuel cell/ battery/ ultracapacitor EV energy consumption optimization. The results from RL EMS showed 7-10% equivalent fuel saving at highway and city driving cycles compared with equivalent consumption minimization strategy. However, aging effect was not considered. In this study, a tabular $Q$-learning algorithm is formulated to minimize the energy consumption and battery degradation. The battery aging model is validated with experimental data and the validation procedure is given. The $Q$-learning EMS is trained on a forward looking EV propulsion plant model, which contains battery, supercapacitor, electric motor, gear, and vehicle dynamics. As the comparison, two heuristic EMS are also proposed and optimized using particle swarm optimization. Overall, the contribution of this paper can be listed as follows:

1) Proposing a tabular $Q$-learning based energy management method for EV to optimize the energy and battery degradation at the same time;

2) Proposing two heuristic energy management methods for EV and the parameters of heuristic methods are optimized by Particle Swarm Optimization algorithm;

3) Providing a comparative study on four energy management methods, which includes a baseline method without ultracapacitor, two heuristic methods with ultracapacitor and a $Q$-learning method with ultracapacitor;

4) Providing a detailed validation procedure for battery aging model and validating the model with experimental data using Genetic Algorithm.

The rest of the paper is organized as follows: Section 2 presents the vehicle propulsion system architecture and plant model. Section 3 introduces the $Q$-learning and two heuristic methods. Then the EMS are integrated into the energy and battery aging minimization problem. Section 4 analyzes the results of $Q$-learning and heuristic methods. Finally, the paper ends with a conclusion and future research.

## 2. Modeling of EV

The EV model includes driver model, propulsion system model and vehicle dynamics. The driver model determines acceleration and braking pedals position based on vehicle speed tracking error. The propulsion system model includes lithium-ion battery, ultracapacitor, and electric motor/ generator (EM). The propulsion system model takes the pedal position command from the driver model and converted the command into torque command, which is then satisfied by the EM torque output. The RL control determines the utilization of lithium-ion battery and ultracapacitor to supply the EM power. The vehicle dynamics model takes the EM torque output, integrates different forces added to the vehicle and calculate the vehicle speed.

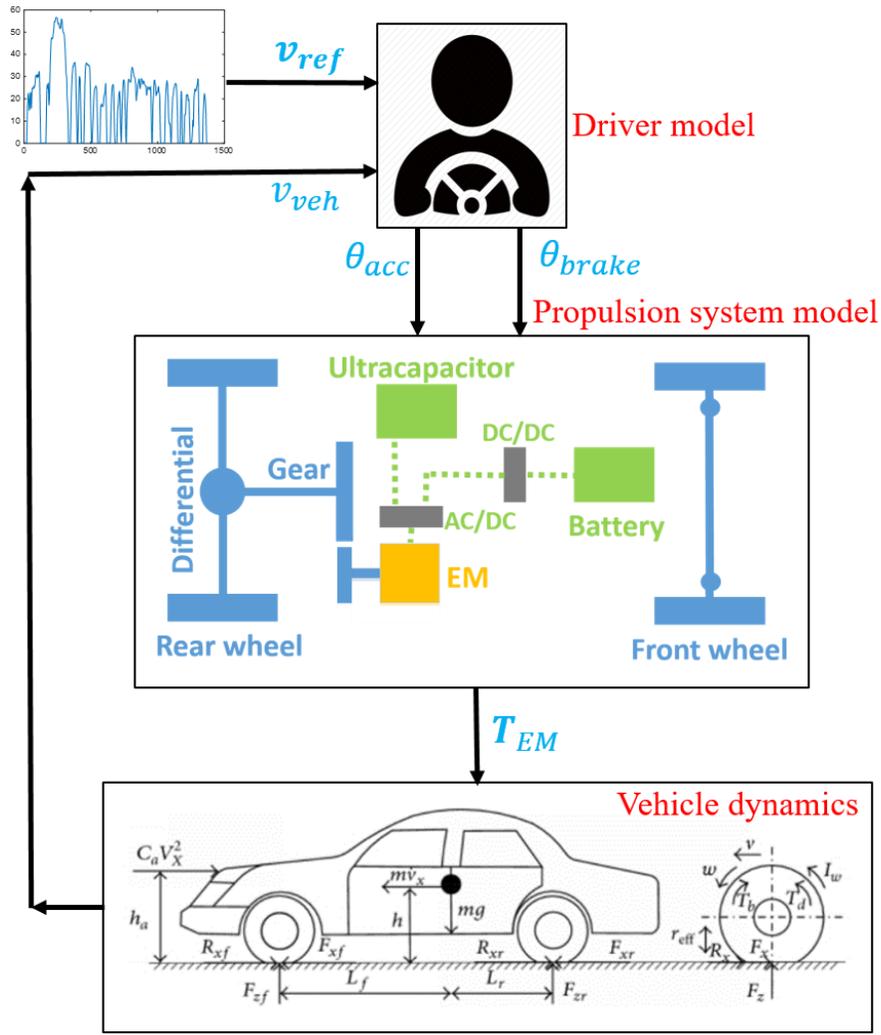

Fig. 1. Vehicle model screenshot in Matlab/Simulink environment.

The propulsion system architecture is shown in the middle of Fig. 1. The EM outputs power to the rear wheel. The EM and rear final differential are connected by a fixed set of gears with the gear ratio 9.59 and the real final differential does not change the torque when transmitting it to the axle.

2.1 *Driver model*

A feedforward and proportional-integral (PI) control represents the driver to press/ release acceleration pedal and braking pedal. The feedforward control is given below:

$$u_{ff} = \begin{cases} \dfrac{T_1}{T_{EM,max}}, & T_1 \geq 0, \\ \left[(1-\sigma) - \dfrac{H_{CG}}{mgr_{whl}B_w}\right]\dfrac{T_1}{-T_{EM,min}}, & T_1 < 0 \end{cases} \quad (1)$$

$$T_1 = \dfrac{\dot{v}J_v}{r_{whl}} + r_{whl}(c_0 + c_1 v + c_2 v^2 + mgsin(\beta)) \quad (2)$$

where $T_{EM,max}$ and $T_{EM,min}$ are the max and min EM torque, $H_{CG}$ is the height of vehicle central gravity and is 500mm, $m$ is the vehicle curb weight, $g$ is the gravity constant, $r_{whl}$ is the wheel radius, $B_w$ is the wheel base and is 2550mm, $J_v$ is the vehicle inertia and is 150kgm2, $c_0, c_1, c_2$ are the road law coefficients and are 105.95, 0.01 and 0.434, respectively, $\beta$ is the road slop. The output of the PI control, acceleration pedal position and braking pedal position are as follows:

$$u_{fb} = k_p e_v(t) + k_i \int_0^t e_v(\tau) d\tau \tag{3}$$

where $e_v$ is vehicle speed tracking error. P/ I gains $k_p, k_i$ are 0.25/ 0.03. The summation of feedforward and feedback control represents the driver input:

$$u_{driver} = u_{ff} + u_{fb} \tag{4}$$

$$\theta_{acc} = \begin{cases} \min(1.0, u_{driver}) * 100\%, if\ u_{driver} > 0 \\ 0, if\ u_{driver} \leq 0 \end{cases} \tag{5}$$

$$\theta_{brak} = \begin{cases} \min(1.0, -u_{driver}) * 100\%, if\ u_{driver} < 0 \\ 0, if\ u_{driver} \geq 0 \end{cases} \tag{6}$$

2.2 *EM model*

The EM is powered by ultracapacitor and battery. It also charges ultracapacitor and battery during regenerative braking, which reduces energy consumption. In this study, the EM efficiency is modeled as a map of EM speed and EM torque. The efficiency map integrates all the losses related to the EM.

2.3 *DC/DC converter model*

DC/DC converter connects the AC/DC converter and the battery. The rated power of DC/DC converter is 30kW and its efficiency is modelled as a map of output power and current [20]. The AC/DC converter efficiency is assumed to be constant at 92%.

2.4 *Battery model*

The battery type is LiFePO4 (LFP). The battery is modelled with equivalent circuit model. The output voltage and the current are calculated as follows:

$$U_{bat} = U_{oc} - I_{bat} R_{bat} \tag{7}$$

$$I_{bat} = \frac{P_{bat}}{U_{bat}} \tag{8}$$

where $U_{bat}$ is the battery terminal voltage, $I_{bat}$ is the battery current, $R_{bat}$ is the battery internal resistance, $P_{bat}$ is the battery power output. The Coulomb counting method is used in the SOC estimation for its popularity and high accuracy for the short-term calculation [21]. The battery SOC at time t is calculated as follows:

$$SOC(t) = SOC(0) - \frac{\int_0^t I_{bat}(\tau)d\tau}{Q_{nom}} \tag{9}$$

where $SOC(0)$ is the initial value of SOC and $Q_{nom}$ is the nominal capacity of the battery. Battery capacity loss is calculated using empirical equations [22]:

$$Q_{loss}(SOC, I_c, T_{bat}, Ah) = \sigma(SOC, I_c, T_{bat})Ah^z \tag{10}$$

$$\sigma(SOC, I_c, T_{bat}) = (\alpha SOC + \beta)exp\left(\frac{-E_a + \delta I_c}{R_g(273.15 + T_{bat})}\right) \tag{11}$$

$$I_c = \frac{|I|}{Q_{nom}} \tag{12}$$

where $\sigma$ is the severity factor, $z, \alpha, \beta, \delta$ are coefficients to be identified, $I_c$ is the battery C-rate, $R_g$ is the universal gas constant 8.3145 [J K$^{-1}$ mol$^{-1}$], $E_a$ is the activation energy equal to 31500 [J mol$^{-1}$], $T_{bat}$ is the battery temperature.

The $z, \alpha, \beta, \delta$ coefficients identification procedure is given in Fig. 2. Three capacity loss dataset are used in the identification. Dataset 1 and 2 are from [23] and dataset 3 are from [24]. The capacity loss and Ah-throughput $(Q_{loss}, Ah)$ of three dataset are plotted as dots in Fig. 3. The respective pairs of SOC, C-rate and batter temperature $(SOC, I_c, T_{bat})$ are given at the first step of Fig. 2**Error! Reference source not found.**. Given the three dataset of $(Q_{loss}, Ah)$, three pairs of $(\sigma, z)$ are identified using the built-in Genetic Algorithm in Matlab with generations set as 2 million. The cost function of the parameter identification is constructed using least square principle. Then, average $z$ values and use the average $z$ value to identify three $\sigma$ values using Eq. (10) and three dataset $(Q_{loss}, Ah)$. Finally, identify one pair of $(\alpha, \beta, \delta)$ using Eq. (11), three pairs of $(SOC, I_c, T_{bat})$ from dataset and three identified $\sigma$ values. After the identification, Eq. (10) and Eq. (11) are converted to be:

$$Q_{loss}(SOC, I_c, T_{bat}, Ah) = (2.0161SOC + 4398.5)exp\left(\frac{-31500 + 112I_c}{8.3145(273.15 + T_{bat})}\right)Ah^{0.5715} \tag{13}$$

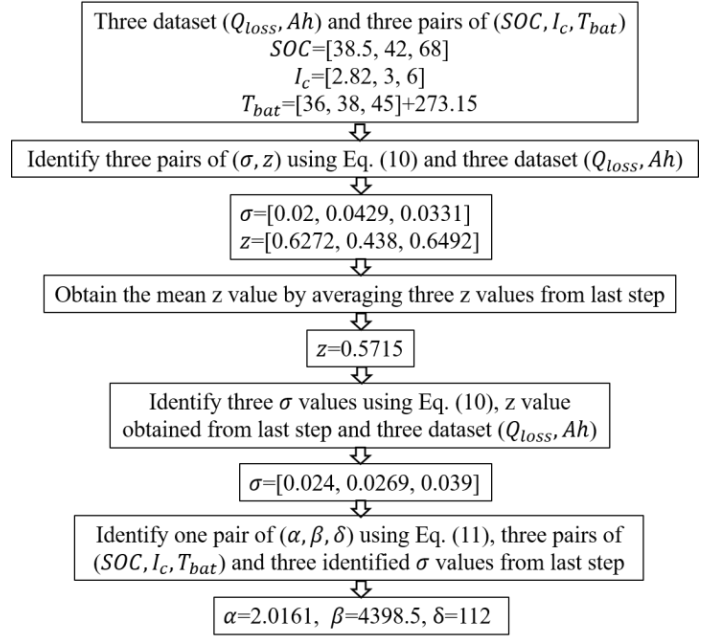

Fig. 2. Parameter identification procedure for severity factor based battery aging model.

Insert the three dataset $(SOC, I_c, T_{bat}, Ah)$ into Eq. (13) and then three smooth capacity loss curves can be obtained, which are plotted against the capacity loss from dataset in Fig. 3. Using Eq. (14), the R squared values for three curves are 0.9085, 0.9458 and 0.9871, respectively.

$$R^2 = 1 - \frac{\sum_{i=1}^{n}(y_{data,i} - y_{model,i})^2}{\sum_{i=1}^{n}(y_{data,i} - y_{data,mean})^2} \qquad (14)$$

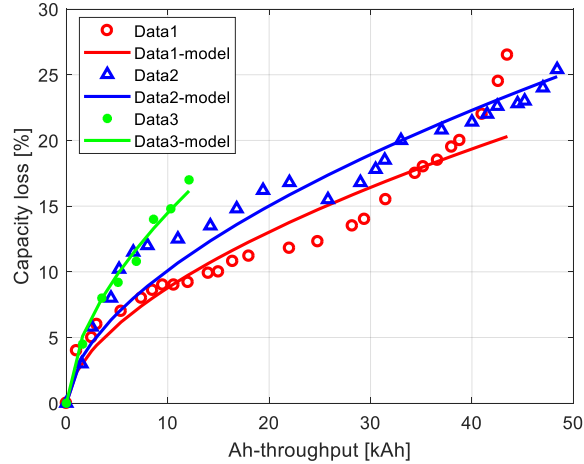

Fig. 3. Comparison of experimental data and the results from identified model.

2.5 *Ultracapacitor model*

The output voltage and the current from the ultracapacitor are calculated as follows [20]:

$$U_{cap,t} = U_{cap,oc} - I_{cap} R_{cap} \qquad (15)$$

$$I_{cap} = \frac{P_{cap}}{U_{cap,t}} \tag{16}$$

where $U_{cap,t}$ is the capacitor terminal voltage, $U_{cap,oc}$ is the open circuit voltage, $I_{cap}$ is the current and $R_{cap}$ is the capacitor resistance, $P_{cap}$ is the power output of capacitor. The current of ultracapacitor can also be expressed as:

$$I_{cap} = C_{cap} U_{cap,max} S\dot{O}V_{cap} \tag{17}$$

where $C_{cap}$ is the capacitance, $U_{cap,max}$ is the maximum voltage at full charge status, $SOV_{cap}$ is the state of voltage. Using Eq. (17), state of charge can be derived as follows:

$$SOV_{cap}(t) = SOV_{cap}(0) - \frac{\int_0^t I_{cap}(\tau)d\tau}{C_{cap} U_{cap,max}} \tag{18}$$

2.6 *Vehicle dynamics model*

The vehicle force balance is as follows:

$$F_{acc} = F_{air} + F_{roll} + F_{grade} + F_{trac} \tag{19}$$

where $F_{acc}$ is the vehicle acceleration force, $F_{air}$ is the aerodynamic resistance force, $F_{roll}$ is the rolling resistance force, $F_{grade}$ is the road grade force, $F_{trac}$ is the vehicle traction force. The five forces are calculated as follows:

$$F_{acc} = ma \tag{20}$$

$$F_{air} = \frac{1}{2}\rho C_d A v_{veh}^2 \tag{21}$$

$$F_{roll} = \cos(\beta) f_{roll} mg \tag{22}$$

$$F_{grade} = \sin(\beta) mg \tag{23}$$

$$F_{trac} = F_{EM} \tag{24}$$

where $m$ is the vehicle weight, $a$ is vehicle acceleration, $\rho$ is the ambient air density, $C_d$ is the air drag coefficient, $A$ is the vehicle front projection area, $v_{veh}$ is the vehicle longitudinal velocity, $\beta$ is the road slop, $f_{roll}$ is the rolling resistance, $F_{EM}$ is the force output by the EM. The EM force is calculated from the wheel torque as follows:

$$F_{EM} = \frac{T_{whl,rear}}{r_{whl}} \tag{25}$$

where $r_{whl}$ is the radius of wheel, the rear wheel torque is calculated as follows:

$$T_{whl,rear} = T_{EM} r_{EM} \tag{26}$$

where $T_{EM}$ is the EM torque output, $r_{EM}$ is the EM reduction gear ratio, $\eta_{EM,mech}$ is the EM mechanical efficiency. Battery power output is a function of discharge and charge status as follows:

$$P_{EM} = P_{bat} + P_{cap} \tag{27}$$

$$P_{EM} = \begin{cases} \dfrac{\omega_{EM} T_{EM}}{\eta_{EM}}, discharge \\ \omega_{EM} T_{EM} \eta_{EM}, charge \end{cases} \tag{28}$$

$$\eta_{EM,elec} = f(\omega_{EM}, T_{EM}) \tag{29}$$

$$\omega_{EM} = \omega_{whl,rear} r_{EM} \tag{30}$$

$$\omega_{whl,rear} = \frac{v_{veh}}{r_{whl}} \tag{31}$$

where $\eta_{EM,elec}$ is the EM electrical efficiency, $\omega_{whl,rear}$ is the rear wheel angular velocity. The vehicle torque demand is satisfied by EM propulsion torque and braking torque as follows:

$$T_{dmd} = T_{prop} + T_{brake} \tag{32}$$

$$T_{prop} = \theta_{acc} T_{ref} \tag{33}$$

$$T_{brake} = \theta_{brake} T_{EM,min} \tag{34}$$

where acceleration and braking pedal positions $\theta_{acc}, \theta_{brake}$ are calculated in Eq. (5) and Eq. (6), respectively.

The specification of the EV is listed in Table 1.

Table 1. Parallel HEV specification.

| Parameters | Values |
| --- | --- |
| Curb weight | 1722 kg |
| EM max torque | 400 Nm |
| EM max power | 143 kW |
| Battery series connection | 98 |
| Battery parallel connection | 60 |
| Battery capacity (single cell) | 2.4 Ah |
| Ultracapacitor number | 50 |
| Ultracapacitor series connection | 1 |
| Ultracapacitor parallel connection | 50 |
| Ultracapacitor capacitance (single unit) | 1200 F |

3. **Supervisory Control Strategies**

3.1 *Q-learning*

*Q*-learning is an adaptive optimal control algorithm [25]. It is adaptive control because its $Q$ table values can be adjusted based on real-time control performance. It is optimal control because it uses Bellman Equation [26] from Dynamic Programming to optimize the collective rewards to the future. The main principle of $Q$-learning algorithm is action selection based on current state and reward feedback from the environment. The $Q$ value update equation is given below:

$$Q(s,a) = (1-\mu)Q(s,a) + \mu\left[R + \gamma \max_a Q(s',a)\right] \tag{35}$$

where $Q$ value is a function of state $s$ and action $a$, $\mu$ is the learning rate, $R$ is the reward obtained in the state transition from $s$ to $s'$, $\gamma$ is the discount factor to penalize the future $Q$ value and pay more attention on the most recent reward. In this study, the two action are the power thresholds of charge $P_{engage,chg}$ and discharge $P_{engage,dischg}$ to engage ultracapacitor as follows:

$$P_{bat} = \begin{cases} P_{EM}, P_{EM} < P_{engage,dischg} \text{ or } P_{EM} > P_{engage,chg} \\ P_{engage,dischg}, P_{EM} \geq P_{engage,dischg} \\ P_{engage,chg}, P_{EM} \leq P_{engage,chg} \end{cases} \tag{36}$$

$$P_{cap} = \begin{cases} 0, P_{EM} < P_{engage,dischg} \text{ or } P_{EM} > P_{engage,chg} \\ P_{EM} - P_{engage,dischg}, P_{EM} \geq P_{engage,dischg} \\ P_{EM} - P_{engage,chg}, P_{EM} \leq P_{engage,chg} \end{cases} \tag{37}$$

The state of this study is the vehicle power demand and ultracapacitor SOC - $SOC_{cap}$. The vehicle power demand equals to the EM power output $P_{EM}$. Each state is discretized into 5 values. The lower and upper boundaries for the states are -30kW/ 50kW for the vehicle power demand and 0%/ 100% for the ultracapacitor SOC. Each action is discretized into 100 values. The lower and upper boundaries for the actions are -20kW/ 0kW for $P_{engage,chg}$ and 0kW/ 40kW for $P_{engage,dischg}$. The state and action discretization analysis can be found in [27]. The reward in this study is the battery severity factor and energy consumption as follows:

$$R = -w_E \frac{E_{bat} + E_{cap}}{E_{bat,norm} + E_{cap,norm}} - (1-w_E)\frac{\sigma}{\sigma_{norm}} + b \tag{38}$$

$$E_{bat} = P_{bat}\Delta t \tag{39}$$

$$E_{cap} = P_{cap}\Delta t \tag{40}$$

where $\sigma$ is calculated in Eq. (11), $w_{aging} = 0.5$ is the aging weight, $E_{bat}$ is the energy consumption by battery, $E_{cap}$ is the energy consumption by ultracapacitor, $b$ is the constant bias term to ensure the reward to be positive, which is set as 1 in this study.

In this study, the schematic of the $Q$-learning contains two block: the black box environment block and $Q$-learning agent block. The vehicle is the in the block of black box environment and the $Q$-learning is in the block of $Q$-learning agent. The real-time information transmitted between these two blocks are the action, state and reward. The reason to call environment black box is that the $Q$-learning agent block only receive state and reward information from the environment and it does not know the rest parameters, which is attributed to the model-free characteristic of $Q$-learning. Vehicle also only takes the action signal from the $Q$-learning agent. $Q$-learning agent updates the $Q$ values following Eq. (35) and information receiving from the environment block. The pseudocode of the proposed $Q$-learning algorithm is given in Table 2. It first initializes $Q$ values and undergoes experience exploration. $\varepsilon$-greedy exploration method is used. During the action determination process, a random action is taken with $\varepsilon$ probability and the greedy action is taken with $(1 - \varepsilon)$ probability. Greedy action is the action pointing to the largest $Q$ value among all the $Q$ values at the given state. After the exploration, the experience is evaluated. The selected experience is used for $Q$ value update. As the learning process continues, the experience evaluation criteria are updated.

Table 2. Pseudocode of $Q$-learning algorithm used in this study.

| $Q$-learning Algorithm |
|---|
| 1 Initialize $Q(s, a)$ with zeros, for all $s \in S, a \in A(s)$. |
|     Initialize $R_{tot}$ with zero. |
| 2 **for** $i \in (1, \ldots, N)$ do (for each episode): |
| 3   **Experience exploration:** |
| 4   Initialize s |
| 5   **for** $j \in (1, \ldots, M)$ **do** (for each time step of episode): |
| 6     Choose action $a_j$ at state $s$ using policy derived from $Q$ ($\varepsilon$-greedy action selection method, $\varepsilon = 0.1$) |
| 7     Take action $a_j$, observe $R_j$, $s_{j+1}$ from environment |
| 8   **end for** |
| 9   **Experience evaluation:** |
| 10   **if** $\sum_1^M R_j > R_{tot}$ **do** |
| 11     $Q$ value **function update:** |
| 12     **for** $j \in (1, \ldots, M)$ **do** (for each time step of episode): |
| 13       $Q(s_j, a_j) = (1 - \mu)Q(s_j, a_j) + \mu \left[R_j + \gamma \max_a Q(s_{j+1}, a)\right]$ |
| 14     **end for** |
| 15     **Experience evaluation criteria update:** |
| 16     Initialize s |
| 17     **for** $j \in (1, \ldots, M)$ **do** (for each time step of episode): |
| 18       Choose action $a_j$ at state $s$ using policy derived from $Q$ ($\varepsilon$-greedy action selection method, $\varepsilon = 0$) |
| 19       Take action $a_j$, observe $R_j$, $s_{j+1}$ from environment |
| 20     **end for** |
| 21     $R_{tot} = \sum_1^M R_j$ |
| 22   **end if** |
| 23 **end for** |

### 3.2 *Heuristic method*

Two heuristic methods are proposed in this study. In the heuristic method 1, ultracapacitor power is a function of ultracapacitor SOV and vehicle power demand $P_{dmd}$. These two parameters are selected as the function input because of their large impact on ultracapacitor operation. The eight constants $a_{i,j}$ will be optimized later in this section.

**Heuristic Method 1:**

$$P_{cap} = \begin{cases} (a_{1,1}SOC_{cap} + a_{1,2}P_{dmd} + a_{1,3}SOC_{cap}P_{dmd} + a_{1,4})P_{dmd}, & (P_{dmd} \geq 0 \text{ and } SOC_{cap} > SOC_{cap,min}) \\ (a_{2,1}SOC_{cap} + a_{2,2}P_{dmd} + a_{2,3}SOC_{cap}P_{dmd} + a_{2,4})P_{dmd}, & (P_{dmd} < 0 \text{ and } SOC_{cap} < SOC_{cap,max}) \\ 0, & else \end{cases} \quad (41)$$

In the heuristic method 2, ultracapacitor is only engaged in the propulsion loop when vehicle power demand reaches certain threshold. In this situation, battery power demand can be attenuated by the ultracapacitor engage and battery aging slows down.

The ultracapacitor power express is shown in Eq. (42). The vehicle power demand thresholds ($a_{dischg}, a_{chg}$) are different in vehicle acceleration and braking. $a_{1,ratio}$ and $a_{2,ratio}$ are the ultracapacitor power versus the vehicle power demand ratios in discharging and charging. These four $a$ values will be optimized later in this section. Ultracapacitor power is zero when the thresholds are not met or the thresholds are met, but $SOC_{cap}$ is out of boundaries.

**Heuristic Method 2:**

$$P_{cap} = \begin{cases} a_{1,ratio}P_{dmd}, & (P_{dmd} > a_{disch} \text{ and } SOV_{cap} > SOV_{cap,min}) \\ a_{2,ratio}P_{dmd}, & (P_{dmd} < a_{chg} \text{ and } SOV_{cap} < SOV_{cap,max}) \\ 0, & else \end{cases} \quad (42)$$

The eight and four $a$ parameters in Eq. (41) and (42) are optimized using particle swarm optimization (PSO) algorithm [28]. PSO is motivated by the animals that hunt in a large group. In the process of prey searching, the predator exchange information frequently. The prey that is spotted by an individual predator will be locked by a large group of predator. The key of PSO is the update of particle velocity and position, which can be expressed as follows:

$$v_i^{k+1} = I^k v_i^k + a_1 c_{1,i}(P_i - x_i^k) + a_2 c_{2,i}(S - x_i^k) \quad (43)$$

$$x_i^{k+1} = x_i^k + v_i^{k+1} \quad (44)$$

where $v$ is the velocity of the particle movement, $i$ is the $i^{th}$ particle, $k$ is the number generation, $I$ is the particle inertia, $a_1, a_2$ are the acceleration constants, $c_1, c_2 \in (0,1)$ are random value, $P_i$ is the optimal position throughout the optimization by i[th] particle, $S$ is the global optimal position throughout the optimization considering all particles. The population and generation sizes are setup as 20 and 20, respectively. The cost function Eq. (45) of the PSO is the same as the reward function Eq. (38) in $Q$-learning except for the bias and the reverse of the signs. The reverse of the signs converts the maximization problem to minimization problem. The optimization results indicate that heuristic methods take around 10 generations to converge.

$$J_{PSO} = w_E \frac{E_{bat} + E_{cap}}{E_{bat,norm} + E_{cap,norm}} + (1 - w_E) \frac{\sigma}{\sigma_{norm}} \quad (45)$$

Table 3. PSO identification results for two heuristic methods.

| Heuristic method 1 | Parameter | $a_{1,1}$ | $a_{1,2}$ | $a_{1,3}$ | $a_{1,4}$ | $a_{2,1}$ | $a_{2,2}$ | $a_{2,3}$ | $a_{2,4}$ |
|---|---|---|---|---|---|---|---|---|---|
| | PSO results | -3.89 | -4.99 | 2.12 | -0.63 | 0.29 | -3.87 | -4.01 | -4.62 |
| Heuristic method 2 | Parameter | $a_{dischg}$ | | $a_{chg}$ | | $a_{1,ratio}$ | | $a_{2,ratio}$ | |
| | PSO results | 34301.00 | | -10210.17 | | 0.83 | | 0.70 | |

3.3 *Baseline without ultracapacitor*

The other baseline control is deployed on the vehicle without ultracapacitor. The energy source is only from battery and there is no need to split vehicle power demand into two power sources. In this baseline strategy, the Eq. (27) is modified and the battery power output equals to the electric motor power as shown in Eq. (46). As the battery is all in during vehicle acceleration and braking, thus there is no need for energy management or optimization.

$$P_{EM} = P_{bat} \tag{46}$$

## 4. Simulation and Results

Urban Dynamometer Driving Schedule (UDDS) driving cycle is used in the $Q$-learning update. The default states and action setup are summarized in Table 4. The boundaries of speed are determined by the driving cycle speed trajectory. The vehicle power demand boundaries are obtained from the vehicle driving cycle simulation. The initial values of entire $Q$ value functions are zeros. In the results section, part of the setup conditions are changed and the changes are specified in the corresponding subsections.

Table 4. $Q$-learning EMS states and action setup.

| Parameters | States | | Action | |
|---|---|---|---|---|
| | Vehicle power demand | Ultracapacitor SOC | Ultracapacitor engage threshold at discharge | Ultracapacitor engage threshold at charge |
| Lower boundary | -30 kW | 0 % | 0 kW | -20 kW |
| Upper boundary | 50kW | 100 % | 40 kW | 0 kW |
| Discretization number | 5 | 5 | 100 | 100 |

### 4.1 Q-learning results analysis

3000 iterations are conducted for the $Q$-learning. The highest sum of rewards converge after 300 iterations and the sum of rewards reduces by 6% compared with initial value. For a better explanation, more variables are shown in Fig. 4. In Fig. 4(a), the vehicle perfectly follows driving cycle speed. Battery SOC drops about 3% in subplot (f) and ultracapacitor SOC fluctuates multiple times between 60%-100% in Fig. 4(d). The SOC fluctuation difference is the result of large capacity difference between two energy storage devices (i.e., given the similar magnitude of power level, large capacity has smaller variation). Comparing Fig. 4(b) and (c), EM power share the shape with battery power except for the negative part. The negative power in Fig. 4(b) represents the vehicle regenerative braking, which is indicated by the deceleration in Fig. 4(a). As shown in Fig. 4(b), most regenerative breaking power is recovered by ultracapacitor. Battery merely has negative power during the braking events. After each braking event, ultracapacitor SOC rises to a high level. At the very next vehicle acceleration (i.e., positive output power in Fig. 4(b)), ultracapacitor outputs power and SOC drops to around 60%. $Q$-learning actions are shown in Fig. 4(e). The ultracapacitor engage power at discharge jumps between 0 and 40 kW, while engage power at charge jumps between 0 and -20kW. Detailed ultracapacitor operation is explained in the coming subsection using the zoom-in window of UDDS simulation.

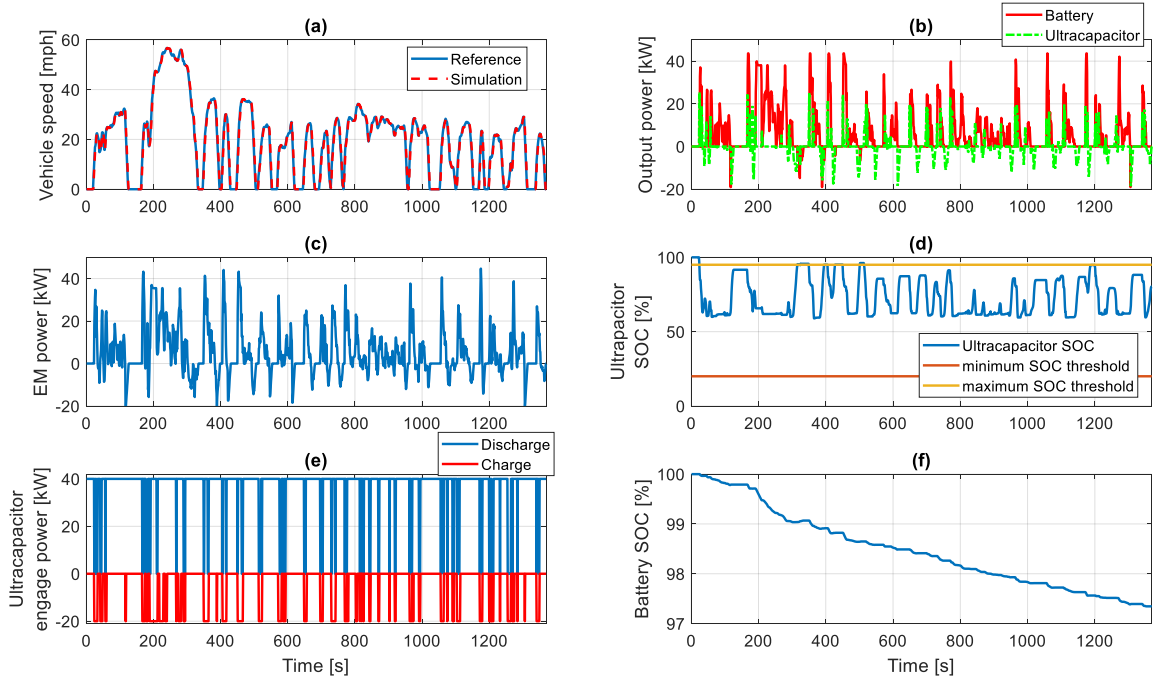

Fig. 4. Vehicle speed, battery/ ultracapacitor power, EM power, ultracapacitor SOC, ultracapacitor engage power and battery SOC using *Q*-learning as the EMS.

Zoom-in window between 400s and 510s of Fig. 4 is shown in Fig. 5. Four ultracapacitor discharge moments are highlighted by the vertical red dash line in the zoom-in window. They are corresponding to the time of 406s, 417s, 451s and 465s. At these four moments, ultracapacitor power shows spike in Fig. 5(b) and its SOC drops as shown in Fig. 5(d). In Fig. 5(e), the discharge ultracapacitor threshold have four valleys at those four moments, which enables the ultracapacitor engage. According to Eq. (37), ultracapacitor power equals to $P_{EM} - P_{engage,dischg}$ when $P_{EM}$ is greater than $P_{engage,dischg}$. At those four valleys, ultracapacitor engage power threshold $P_{engage,dischg}$ is 0kW. Thus the ultracapacitor power equals to the EM power $P_{EM}$. The ultracapacitor engage power threshold for the four moments can be clearly observed in the optimal policy map as shown in Fig. 6(g). The first and second moments are crossed by the green line between 400s and 418s. During the 18s period, vehicle power demand increases from 0kW to 40kW and then drops to 0kW. Ultracapacitor SOC drops from 95% to 75%. The trajectory of the green line starts from 40kW, drops to 0kW, rises to 40kW, drops to 0kW and finally goes up to 40kW, which explains the two valleys in Fig. 5(e). Similar trajectory is observed between 440s and 468s in Fig. 5(g), which explains the two valleys during that period in Fig. 5(e).

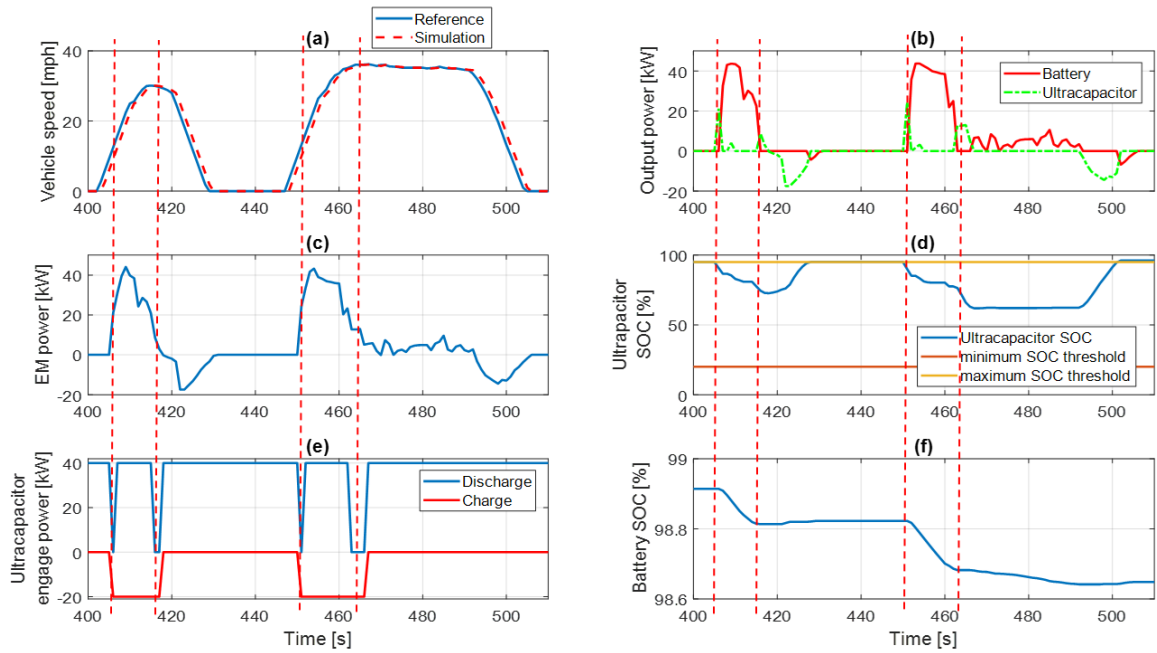

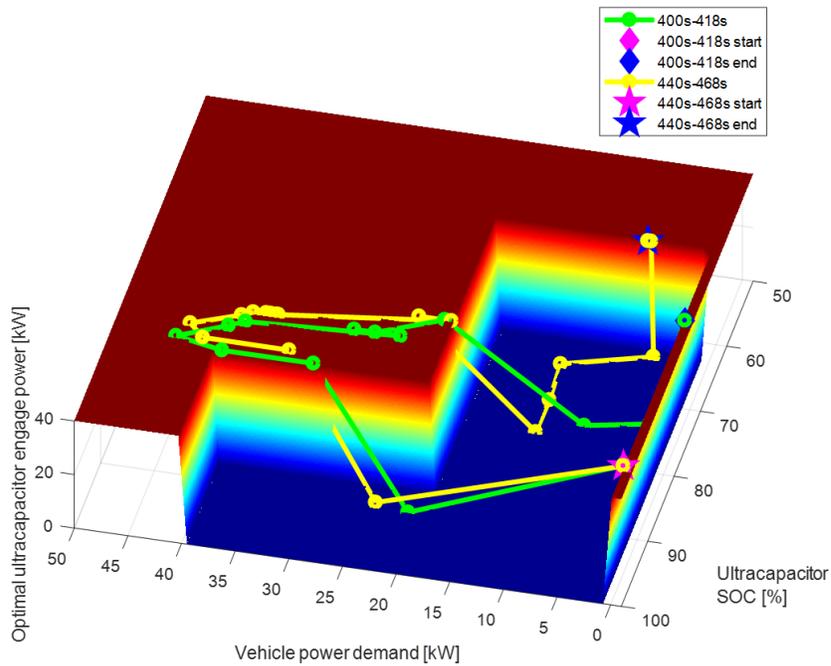

Fig. 5. Vehicle operation variables using *Q*-learning as the EMS. Four battery and ultracapacitor discharge scenarios are highlighted in vehicle operation variable trajectories.

The ultracapacitor charge situation can be explained in a similar way. Two charge moments are highlighted by vertical red dahs lines in Fig. 6. At those two moments, vehicle speeds are dropping as shown in Fig. 6(a). The negative EM power at those two moments represent the vehicle regenerative braking. Instead of battery, ultracapacitor shows negative power at those two moments and stores the generated braking energy as shown in Fig. 6(b). It is observed in Fig. 6(d) that the ultracapacitor SOC rises from 75% to 95% at the first braking event and rises from 65% to 95% at the second braking event. The ultracapacitor engage power stay at 0 at those two braking events. According to Eq. (37), ultracapacitor charge power equals to EM power $P_{EM}$ when $P_{EM}$ is less than $P_{engage,chg}$. At the two braking events, the trajectories of ultracapacitor engage power $P_{engage,chg}$ are shown in Fig. 6(g). The green line indicates the $P_{engage,chg}$ trajectory between 420s and 428s. The yellow line indicates the $P_{engage,chg}$ between 493s and 508s. Even though the ultracapacitor SOC rises and vehicle power demand varies, the two trajectories lay on the 0kW surface, forcing the ultracapacitor power equal to the EM power and charge.

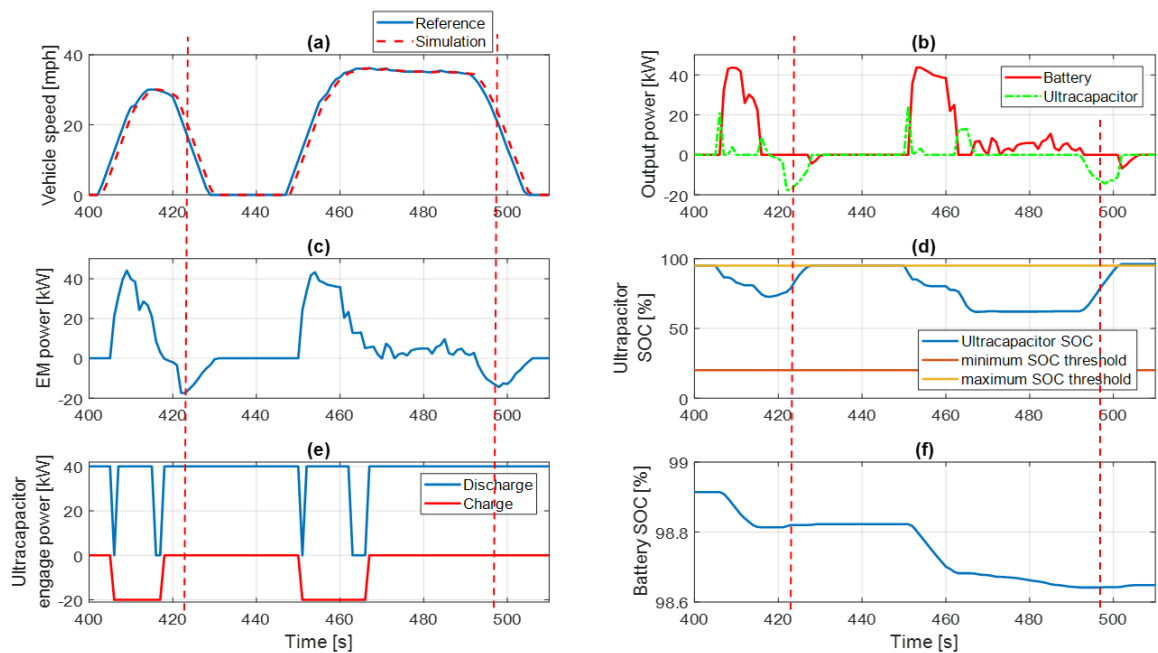

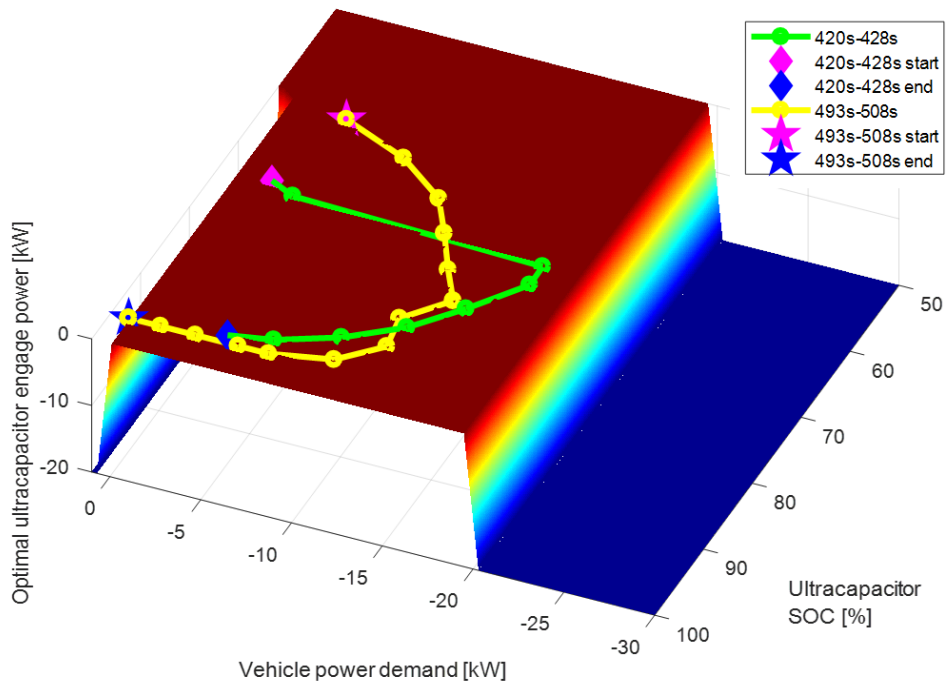

Fig. 6. Vehicle operation variables using *Q*-learning as the EMS. Two battery and ultracapacitor charge scenarios are highlighted in vehicle operation variable trajectories.

4.2 *EMS comparison*

The vehicle simulation with four EMS methods (i.e., Q-learning, heuristic method 1, heuristic method 2 and baseline with no ultracapacitor) are conducted over a UDDS driving cycle. The vehicle speed, EM power, battery power, ultracapacitor power, battery SOC and ultracapacitor SOC are shown in Fig. 7. Over the entire cycle, one observation is that the vehicle without ultracapacitor charges battery at regenerative braking as shown in the green line of the negative battery power in Fig. 7(c). Heuristic method 2 is spotted charging battery as well at multiple regenerative braking events with lower charging power. The ultracapacitor power difference among the three strategies is not very obvious in the entire driving cycle span. To get more details, the zoom-in window between 400s and 510s are shown on the right side of Fig. 7. EM of heuristic method 1 outputs power after the rest three methods at two vehicle acceleration events (i.e., 408s and 452s). That is because ultracapacitor of heuristic method 1 power output at those two events maintains the longest duration among three methods with ultracapacitor as shown in the zoom-in window of Fig. 7(d). That long power output of heuristic method 1 is compensated by the extended ultracapacitor charge at 428s and 503s. The long time discharge and charge are explained by larger SOC change in Fig. 7(f). Another difference among the four methods is that the battery power of Q-learning retracts and ultracapacitor steps at the vehicle peak speed – 416s and 463s in the zoom-in window of Fig. 7(c) and (d), which reduces the battery usage. According to Fig. 7(f), heuristic method 1 and Q-learning share ultracapacitor SOC variation trajectory with different magnitude. Heuristic method 2 has less ultracapacitor SOC variation. At the end of the driving cycle, battery SOC differs among four methods. The left SOC for Q-learning, no ultracapacitor, heuristic method 2 and heuristic method 1 are 97.35%, 97.27%, 97.25%, and 97.15%, respectively. As one driving cycle is too short to investigate the energy efficiency. The range simulation and battery aging simulation are conducted with many more repeated driving cycles in the coming subsection.

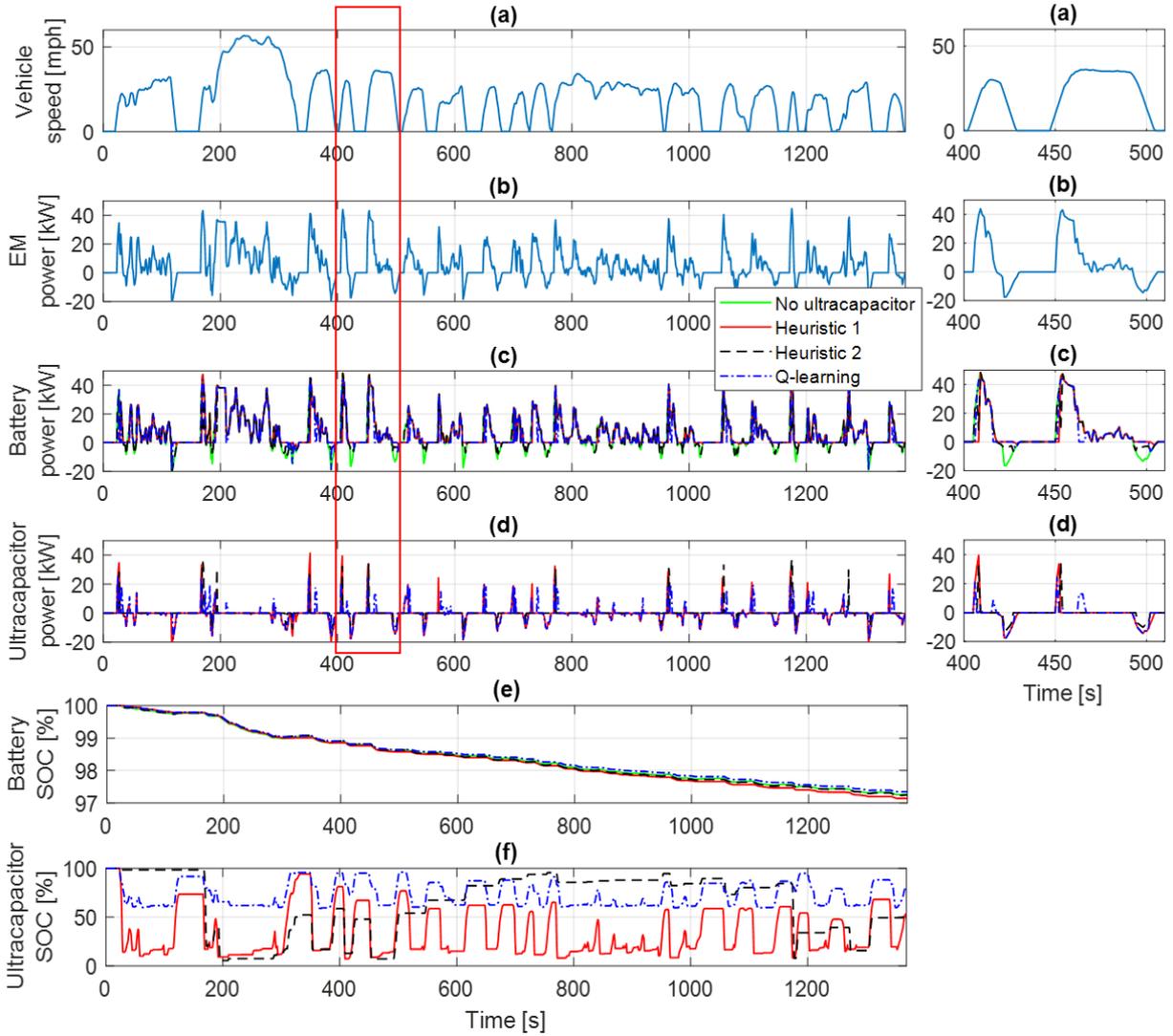

Fig. 7. Comparison of the proposed Q-learning method and the rest three methods: (a) vehicle speed, (b) EM power, (c) battery power, (d) ultracapacitor power, (e) battery SOC, and (f) ultracapacitor SOC.

To compare the energy efficiency and battery aging effect, repeated driving cycles are simulated with the four different methods. Worldwide Harmonised Light Vehicle Test Procedure (WLTP) driving cycle [29] is used to validate the four methods. For the range simulation, the battery initial SOC is set as 100% and vehicle repeats the driving cycle until battery SOC reaches 0.1%. For the 500 cycles capacity loss simulation, initial battery capacity loss is 0 and the vehicle repeats 500 driving cycles. Once battery SOC reaches 0.1%, it is then fully charged to 100% and continues the rest driving cycles. The results are summarized in Table 5. At the repeated UDDS driving cycles, the vehicle with Q-learning has the longest range – 251.8 miles, which is longer than the vehicle without ultracapacitor. However, the vehicles equipped with ultracapacitor using heuristic method 1 and 2 have shorter range than the vehicle without ultracapacitor. This indicates that the ultracapacitor has certain ability to increase the electric vehicle range if it is controlled well. The range matches the left SOC in Fig. 7(e). The 500 UDDS cycles capacity loss show greater difference than the range. The vehicle with Q-learning has 0.34% capacity loss, while the vehicle without ultracapacitor posts 0.41% capacity loss, which is 20.6% increase. The vehicle simulations using heuristic method 1 and 2 show certain capacity loss reduction when compared with the vehicle simulation results without ultracapacitor. The capacity loss reduction by the Q-learning, heuristic method 1 and heuristic method 2 can be explained by the reduction of Ah-throughput in Table 5. The Q-learning shows the least Ah-throughput among the four methods. The difference and order among the four methods are the same between Ah-throughput and capacity loss. According to Eq. (10), the capacity loss is proportional to the Ah-throughput with the power of 0.5715, given the similar severity factor among the four methods.

Table 5. The range, 500 driving cycle capacity loss and Ah-throughput comparison among four methods.

| Cycles | Variables | No ultracapacitor | Heuristic method 1 | Heuristic method 2 | Q-learning |
|---|---|---|---|---|---|
| UDDS | Range [miles] | 246.7 | 232.9 | 240.5 | 251.8 |
|  | 500 cycles capacity loss [%] | 0.41 | 0.35 | 0.39 | 0.34 |
|  | Ah-throughput [Ah] | 85.8 | 63.2 | 77.6 | 61.5 |
| WLTP | Range [miles] | 213.5 | 206.8 | 209.8 | 216.8 |
|  | 500 cycles capacity loss [%] | 0.61 | 0.54 | 0.58 | 0.54 |
|  | Ah-throughput [Ah] | 168.9 | 138.1 | 156.2 | 137.2 |

## 5. Conclusion

This paper proposes a *Q*-learning based energy management strategy to minimize the battery degradation and energy consumption for a battery/ ultracapacitor powered electric vehicle. In the vehicle propulsion modeling section, battery capacity loss model validation process is described. In the results section, *Q*-learning takes around 300 iterations to converge. To evaluate the *Q*-learning method, two heuristic energy management strategies are proposed and optimized. The heuristic methods optimization process is described in detail. In addition, a baseline vehicle without ultracapacitor is compared with the vehicle using aforementioned three energy management strategies (i.e., one *Q*-learning and two heuristic methods). The *Q*-learning is developed in UDDS driving cycle and validated in WLTP driving cycle. In both driving cycles, *Q*-learning outperforms the rest three strategies both in vehicle range and battery capacity loss. Compared with the baseline vehicle without ultracapacitor, the vehicle using *Q*-learning reduces battery capacity loss by 13-20% and extends the range by 1.5-2%. Compared with the heuristic method 1, *Q*-learning has similar battery capacity loss, but extends range by 4.8-8.3%. Compared with the heuristic method 2, *Q*-learning reduces battery capacity loss by 7.4-14.7%, and extends range by 3.3-5%.

In future studies, few limitations of the presented work will be addressed to improve the presented work. The experiments can validate the proposed energy management method, where the measurement errors should be considered. Moreover, the component sizing of ultracapacitor and battery should be conducted. The efficacy of the proposed energy management strategy on different component sizes need to be investigated. Finally, the proposed method can be improved by considering approximation based *Q*-learning for smoother Q value functions rather than table-based.